# SQUID magnetometer based on Grooved Dayem nanobridges and a flux transformer

E. Trabaldo, C. Pfeiffer, E. Andersson, R. Arpaia, A. Kalaboukhov, D. Winkler, F. Lombardi, and T. Bauch

Department of Microtechnology and Nanoscience
Chalmers University of Technology
SE-41296 Göteborg, Sweden
thilo.bauch@chalmers.se

*Abstract*—We report noise measurements performed on a SQUID magnetometer implementing Grooved Dayem nanobridge of YBCO as weak-links. The SQUID shows magnetic flux noise as low as 10 $\mu\Phi_0/\text{Hz}^{0.5}$. The magnetometer is realized by coupling the SQUID to a flux transformer with a two-level coupling scheme using a flip-chip approach. This improves the effective area of the SQUID and result in a magnetic field noise of 50 fT/Hz$^{0.5}$ at T=77 K.

## I. INTRODUCTION

The growing demand for quantum limited devices has been the main driving force for the realization of superconducting electronics beyond the state-of-the-art. One of the most prominent application of superconductors are magnetic flux and field sensors based on SQUIDs. These devices are already used in many applications, yet, a great amount of effort is put into the improvement of their performance and fabrication procedure. In fact, the main focus has been to achieve high quality High critical Temperature Superconductors (HTS) Josephson Junctions (JJs), the key ingredient of a SQUID, during the last few decades. This has proven to be challenging for cuprate HTS materials, due to their chemical instability and small superconducting coherence length (≈2 nm in the ab-planes).

Different JJ fabrication techniques have been successfully developed for HTS SQUIDs so far. For example, high sensitivity HTS SQUIDs have been realized using grain boundary based JJs [1][2]. These have the disadvantage of a rather complex fabrication procedure, which might limit their use in technological applications. Alternatively, HTS nanoSQUIDs realized with Dayem bridges have also shown low magnetic flux noise properties, in combination with a simplified fabrication procedure [3], compared to grain boundary-based devices. However, the rather large parasitic inductance of Dayem bridges limits their implementation in SQUID magnetometers at 77 K [4][5]. To address the issues of both grain boundary based JJs and Dayem bridges, we propose an alternative weak link choice for SQUIDs.

We present a novel fabrication process of HTS weak links: the nanoscale Grooved Dayem Bridge (GDB) [6], which exhibits Josephson Junction-like behavior. Here, the layout of the bridge and the weak link inside the bridge are realized during one single lithography process on a YBa$_2$Cu$_3$O$_{7-\delta}$ (YBCO) film grown on a single crystal substrate. This results in high-quality weak links with I$_C$R$_N$ products as high as 350 µV [6] and differential resistances much larger than in bare Dayem bridges at T = 77 K. Moreover, such weak links can be defined anywhere on the chip and freely oriented within the film plane. This greatly simplifies the fabrication procedure compared to grain boundary based JJs. GDB based SQUIDs combine the nanofabrication advantages and the device reproducibility, which are typical of Dayem bridges, with the performances, e.g. the magnetic sensitivity, of state-of-the-art SQUIDs based on grain boundary JJs. By combining such GDB based SQUIDs with a YBCO flux transformer realized on a 10×10 mm$^2$ substrate using a flip chip approach [4] we can further improve the magnetic field sensitivity of GDB SQUID magnetometers compared to single layer GDB SQUID

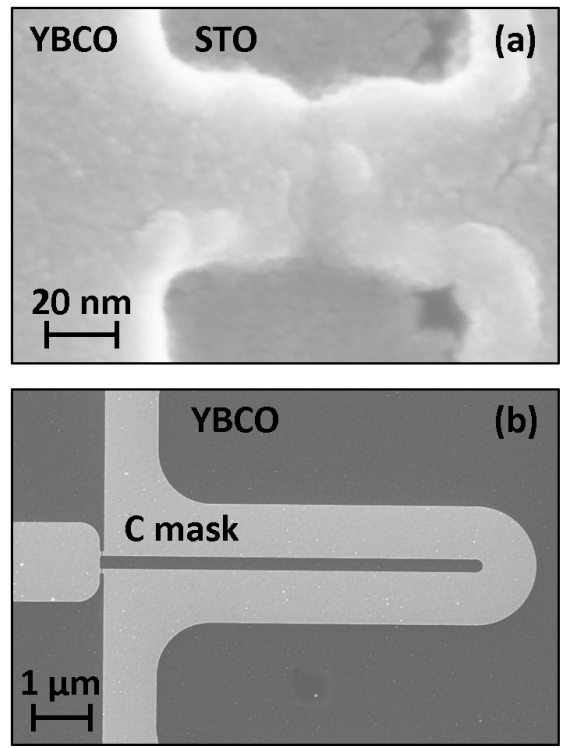

Fig. 1. *(a) SEM image of a GDB, the central part of the nanowire shows the groove. (b) SEM image of the SQUID design with a hairpin loop.*

## II. RESULTS AND DISCUSSION

To obtain GDB-based SQUIDs a thin film of YBCO (50 nm) is deposited by Pulsed Laser Deposition (PLD) on a SrTiO$_3$ (001) substrate. The YBCO is covered with a layer of amorphous carbon of thickness $t_{mask}$=120 nm, also deposited by PLD. Electron beam lithography is used to define the etching mask, which is then transferred into the carbon layer. The latter serves as a hard mask for the YBCO during the final Ar ion beam milling of the YBCO film. This fabrication technique has been used to obtain YBCO nanostructures with pristine quality and lateral dimension down to 65 nm [7][8]. To obtain a GDB, a narrow gap is opened in the nanowire mask, as described in detail in [6]. When the gap in the mask is in the order of $t_{mask}$/2 or smaller, the etching rate of the material inside the gap is reduced compared to other areas of the sample. The final result is a nanobridge with a groove etched along its width, which acts as a weak link. An example of a GDB is shown in Fig.1. Here it can be seen that the fabrication process results in a nanowire with a grooved profile. The final devices show lower critical current $I_c$ and higher differential resistance $\delta V/\delta I$ at T=77 K compared to standard nanowires. A more detailed description of the etching dynamic of GDB is reported in [6]. The result presented here have been obtained on a GDB-based SQUID implementing 200 nm wide GDBs with a gap length of 50 nm.

SQUID magnetometers based on GDB have been previously realized, showing the advantages of this technology. In [6][9], the magnetometers were fabricated with a galvanically coupled pick-up loop to increase the effective area ($A_{eff}$) of the SQUID. Here, to further improve $A_{eff}$, the SQUID is inductively coupled to an external superconducting flux-transformer using a flip-chip setup. The flip-chip approach has been studied in detail in [4] for nanowire-based SQUIDs. In this work we use the same two-level coupling scheme where the SQUID is first galvanically coupled to a washer type pick-up loop, whose design is depicted in Fig.2. The washer is then inductively coupled to the flux-transformer input coil, the size of which matches the size of the inner hole of the washer loop.

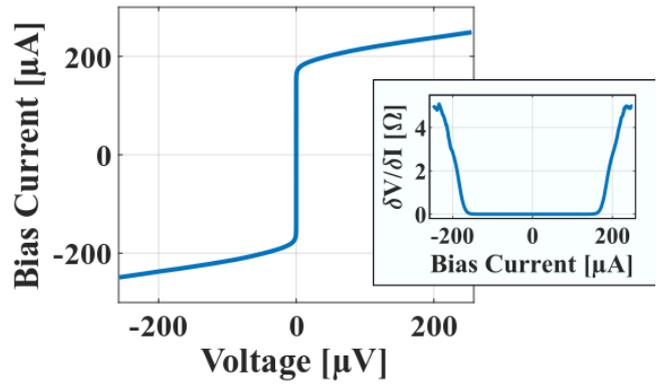

Fig. 2. *Current voltage characteristic (IVC) measured for the SQUID. Inset: Differential resistance for the same device.*

As shown in [4], the effective area of a flip-chip SQUID magnetometers can be approximated as:

$$A_{eff} \approx \frac{\alpha}{2} \cdot \frac{L_C}{\sqrt{L_{pk} \cdot L_{in}}} \cdot A_{eff,FT}$$

Where $A_{eff}$ and $A_{eff,FT}$ are respectively the total effective area of the SQUID and the effective area of the bare flux transformer. $L_C$ is the coupling inductance, given by the inductance of the SQUID loop, $L_{pk}$ is the inductance of the pick-up loop and $L_{in}$ is the inductance of the flux transformer input coil, $\alpha$ is the coupling coefficient between washer and flux transformer. To increase the coupling inductance, hence improve the value of $A_{eff}$, the SQUID is patterned with a hairpin loop, as shown in Fig.1(b). The SQUID loop slit is 12 x0.5 µm² which gives an inductance of 96 pH. The washer loop has an outer lateral dimension of 400 µm and inner diameter of 60 µm. This matches the flux transformer input coil dimensions used in this work, which are the same as the ones used in [4]. The measured effective area of the flip-chip magnetometer is $A_{eff}$ = 0.41 mm². This is consistent with previously reported values, measured on devices using the same flip-chip technology. The $A_{eff}$ obtained here is almost four times larger than what has been achieved for single layer GDB based SQUIDs galvanically coupled to a pick-up loop [6].

The device is cooled down to 77 K in a magnetically shielded room. The Current Voltage Characteristic (IVC) of the device is shown in Fig.3. To compare the weak link performance with previous results, we can define $\delta R$ as the value of differential resistance measured at the working point of the SQUID, i.e. the bias current value at which the voltage modulations are maximized. For this device, $\delta R$=2 Ω, which gives a product $I_c\delta R$=350 µV, which is higher than the value reported previously for GDB SQUIDs [6].

The SQUID is operated in a flux-locked feedback loop (FLL) using a Magnicon SEL-1 SQUID electronics [10]. The device is operated in a bias reversal mode at 40 kHz to reduce the low frequency critical current noise contributions. The measured magnetic flux noise is $S^{1/2}_\Phi$=10 µ$\Phi_0$/Hz$^{0.5}$ above 100

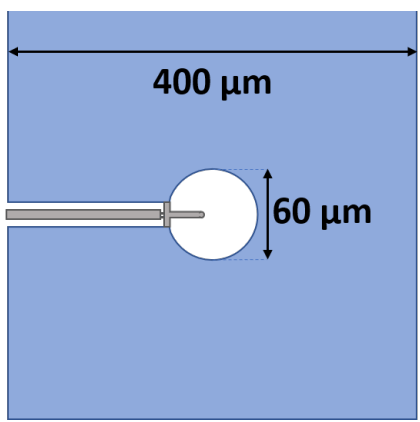

Fig. 3. *Washer type pick-up loop design. The hairpin loop has a length of 12µm.*

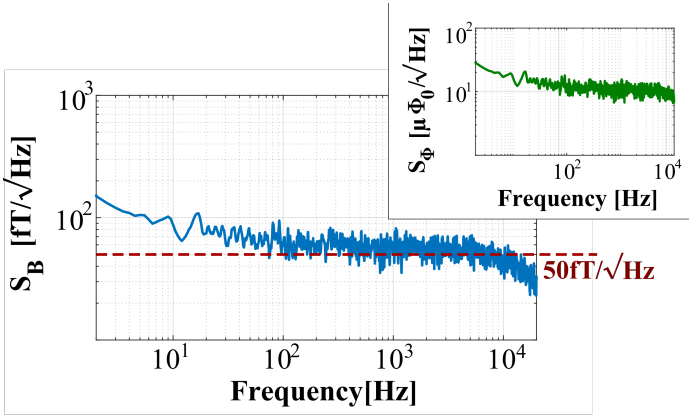

Fig. 4. *Magnetic field noise $S_B$ measured with the flux transformer and bias reversal scheme. The red line indicate the 50 fT/Hz$^{0.5}$ limit. Inset: Corresponding magnetic flux noise $S_\Phi$.*

Hz (see inset of Fig.4). This corresponds to a magnetic field noise $S^{1/2}_B$=50 fT/Hz$^{0.5}$ (see Fig.4). The magnetic flux noise of the device is comparable with the state-of-the-art grain boundary based YBCO SQUIDs, $S^{1/2}_\Phi$=4.5-10 µ$\Phi_0$/Hz$^{0.5}$ [1][11][12][13][14]. The magnetic field noise is lower than the one reported on YBCO nanowire based SQUIDs with flip chip technology [4] and lower than the noise obtained for single layer GDB based SQUID magnetometers [6]. The noise is still an order of magnitude higher than the lowest reported value for flip-chip magnetometers [15], where $S^{1/2}_B$=4 fT/Hz$^{0.5}$, but is comparable with the value reported for single layer grain boundary based SQUIDs coupled galvanically to a pick-up loop, where $S^{1/2}_B$=30-50 fT/Hz$^{0.5}$ [1][11][13].

### III. CONCLUSION

We have used YBCO GDBs as novel nanoscale building blocks in an HTS SQUID magnetometer coupled to an in-plane washer type pick-up loop and an external flux transformer realized on a 10×10 mm$^2$ substrate. The magnetometer has been characterized via transport and noise measurements at T=77 K. The device exhibits low values of white magnetic flux noise, 10 µ$\Phi_0$/Hz$^{0.5}$ and corresponding magnetic field noise, 50 fT/Hz$^{0.5}$. This value is comparable with single layer state-of-the-art magnetometers based on grain boundary JJs [1][11][13]. Further improvement of the field sensitivity can be achieved by implementing a single-level coupling of the flux transformer to the SQUID loop. Here, the GDBs are integrated directly in a single loop washer SQUID, which is then inductively coupled to a flux transformer. This should result in magnetometers with $A_{eff} \approx 2.2$ mm$^2$ [16], corresponding magnetic field noise, 12 fT/Hz$^{0.5}$.

Our work proves the feasibility of Grooved Dayem Bridges as a valid alternative to state-of-the-art JJs for SQUID applications. Here, the development of low noise HTS SQUIDs is crucial not only for technological applications, but also for fundamental research. Moreover, GDBs could open the way to a range of future applications, such as HTS rapid single flux quantum (RSFQ) circuits [17] and high-performance high-frequency HTS superconducting quantum interference filters (SQIFs) [18]. The physics governing this novel type of weak links is not yet fully understood and will require further studies. The electrical transport properties show features which are different from standard Dayem bridges, yet they cannot be fully described by the Resistively Shunted Junction model.

### ACKNOWLEDGMENT

We thank Justin Schneiderman for helpful discussion. This work was been supported in part by the Knut and Alice Wallenberg Foundation (KAW) and in part by the Swedish Research Council (VR). This project has received funding from the ATTRACT project funded by the EC under Grant Agreement 777222. R. A. is supported by the Swedish Research Council (VR) under the project "Evolution of nanoscale charge order in superconducting YBCO nanostructures".

### REFERENCES

[1] Chukharkin, M.; Kalabukhov, A.; Schneiderman, J. F.; Öisjöen, F.; Jönsson, M.; Xie, M.; Snigirev, O. V.; Winkler, D. IEEE Trans. Appl. Supercond. 2013,23

[2] Mitchell, E.; Foley, C. Supercond. Sci. Technol. 2010, 23, 065007.

[3] Arpaia, R.; Arzeo, M.; Baghdadi, R.; Trabaldo, E.; Lombardi, F.; Bauch, T. Supercond. Sci. Technol. 2016, 30, 014008.

[4] Xie, M.; Chukharkin, M.; Ruffieux, S.; Schneiderman, J.; Kalabukhov, A.; Arzeo, M.; Bauch, T.; Lombardi, F.; Winkler, D. Supercond. Sci. Tech. 2017, 30, 115014.

[5] Trabaldo, E., Arpaia, R., Arzeo, M., Andersson, E., Golubev, D., Lombardi, F., & Bauch, T. (2019). Transport and noise properties of YBCO nanowire based nanoSQUIDs. *Superconductor Science and Technology*, *32*(7), 073001.

[6] Edoardo Trabaldo, Christoph Pfeiffer, Eric Andersson, Riccardo Arpaia, Alexei Kalaboukhov, Dag Winkler, Floriana Lombardi, and Thilo Bauch, Nano Letters 2019 19 (3), 1902-1907.

[7] Nawaz, S.; Arpaia, R.; Bauch, T.; Lombardi, F. Approaching the theoretical depairing current in YBa2Cu3O7- x nanowires. Physica C Supercond 2013, 495, 33-38.

[8] Trabaldo, E.; Arzeo, M.; Arpaia, R.; Baghdadi, R.; Andersson, E.; Lombardi, F.; Bauch, T. Noise properties of YBCO Nanostructures. IEEE Trans. Appl. Supercond. 2017, 27 .

[9] Arzeo, M.; Arpaia, R.; Baghdadi, R.; Lombardi, F.; Bauch, T. Toward ultra-high magnetic field sensitivity YBCO nanowire based superconducting quantum interference devices. J. Appl. Phys. 2016, 119, 174501.

[10] Magnicon, SQUID electronics SEL-1, http://www.magnicon.com.

[11] Öisjöen, F.; Schneiderman, J. F.; Figueras, G.; Chukharkin, M.; Kalabukhov, A.; Hedström, A.; Elam, M.; Winkler, D. High-T c superconducting quantum interference device recordings of spontaneous brain activity: Towards high-T c magnetoencephalography. Appl. Phys. Lett. 2012, 100, 132601.

[12] Faley, M.; Poppe, U.; Dunin-Borkowski, R.; Schiek, M.; Boers, F.; Chocholacs, H.; Dammers, J.; Eich, E.; Shah, N.; Ermakov, A. High-Tc DC SQUIDS for magnetoencephalography. IEEE Trans. Appl. Supercond. 2013, 23, 1600705-1600705.

[13] Faley, M.; Meertens, D.; Poppe, U.; Dunin-Borkowski, R. Graphoepitaxial high-Tc SQUIDs. Journal of Physics: Conference Series. 2014; p 042009.

[14] Mitchell, E.; Foley, C. YBCO step-edge junctions with high IcRn. Supercond. Sci. Technol. 2010, 23, 065007.

[15] Faley, M.; Dammers, J.; Maslennikov, Y.; Schneiderman, J.; Winkler, D.; Koshelets, V.; Shah, N.; Dunin-Borkowski, R. High-Tc SQUID biomagnetometers. Supercond. Sci. Technol. 2017, 30, 083001.

[16] Minshu Xie, "Development of High-Tc SQUID magnetometers for on-scalp MEG", PhD diss., Chalmers University of Technology,

[17] Wolf, T.; Bergeal, N.; Lesueur, J.; Fourie, C. J.; Faini, G.; Ulysse, C.; Febvre, P. IEEE Transactions on applied superconductivity 2013, 23, 1101205–1101205.

[18] Mitchell, E.; Hannam, K.; Lazar, J.; Leslie, K.; Lewis, C.; Grancea, A.; Keenan, S.; Lam, S.; Foley, C. Supercond. Sci. Technol. 2016, 29, 06LT